\documentclass[draft]{article}
\usepackage{graphicx}
\title{Auxetic properties and anisotropy of elastic material constants of 2D crystalline media}
\author{Cz. Jasiukiewicz\and T. Paszkiewicz\and S. Wolski\\
\\
Faculty of Mathematics and Applied Physics,\\
 Rzesz\'{o}w University of Technology,\\ 
 Al. Powsta\'{n}c\'{o}w. Warszawy 6,\\
PL-35-959 Rzesz\'{o}w Poland
}
\date{}

\begin{document}
\maketitle

\begin{abstract}
Anisotropies of Young's modulus $E$, the shear modulus $G$, and Poisson's ratio $\nu$ of all 2D symmetry systems are studied. The shear modulus and  Poisson's ratio of 2D crystals have fourfold symmetry. Simple necessary and sufficient conditions on their elastic compliances are derived to identify if any of these crystals is completely auxetic, non-auxetic or auxetic. Examples of all types of auxetic properties of crystals of oblique and rectangular symmetry are presented. Particular attention is paid to 2D crystals of quadratic symmetry. All mechanically stable quadratic crystals are characterized by three parameters belonging to a prism with the stability triangle in the base. Regions in the stability triangle in which quadratic materials are completely auxetic, non-auxetic, and auxetic are established. 
\end{abstract}

\section{Introduction} 
\label{sc:introd}
Mechanical properties of elastic media are described by the bulk, Young's, and the shear modules. Poisson's ratio is another important characteristics of such materials. With the exception of bulk modulus, generally, these characteristics are anisotropic, i.e. they depend on direction $\textbf{n}$ of load and direction $\textbf{m}$ of lateral strain. The Poisson ratio is especially interesting because it can be negative. Materials which exhibit a negative Poisson's ratio are referred to as auxetic media. Auxetics respond to imposed uniaxial tension with lateral extension in place of expected contraction. Such media can find interesting applications in future technologies. Ting and Barnett introduced classification of the auxetic behavior of anisotropic linear media \cite{ting}. 

In our previous papers we derived expressions for the inverse of Young's modulus $E(\textbf{n})$, the inverse of the shear modulus $G(\textbf{m},\textbf{n})$ and Poisson's ratio $\nu(\textbf{m},\textbf{n})$ of all 2D \cite{japawol} and of 3D crystalline media of high and medium symmetry \cite{pawol1}. In another paper \cite{pasz-wol} we studied the above mechanical properties of all mechanically stable cubic media. To achieve this goal, we used a set parameters introduced by Every \cite{every}. All stable cubic elastic materials are represented by points of a prism with the stability triangle in the base. We established regions in the stability triangle in which 3D cubic elastic materials, in accord to Ting and Barnett classification \cite{ting}, are completely auxetic, non-auxetic, and auxetic. 

The aim of the present paper is to advance the study of anisotropy properties of the mentioned material constants of 2D crystals and formulate necessary and sufficient conditions on elastic compliances to identify whether any given 2D material is  completely auxetic, auxetic or non-auxetic. The quadratic materials will be studied in detail. We will show that their elastic properties can be studied in the same manner as those of 3D cubic materials. For this purpose we shall use parameters, which were 2D analogues of parameters introduced by Every  \cite{every} (cf. Appendix \ref{sc:phase-vel}). 

\begin{figure}[h]
	\centering
		\includegraphics[width=11.5cm, bb=104 628 512 792,clip]{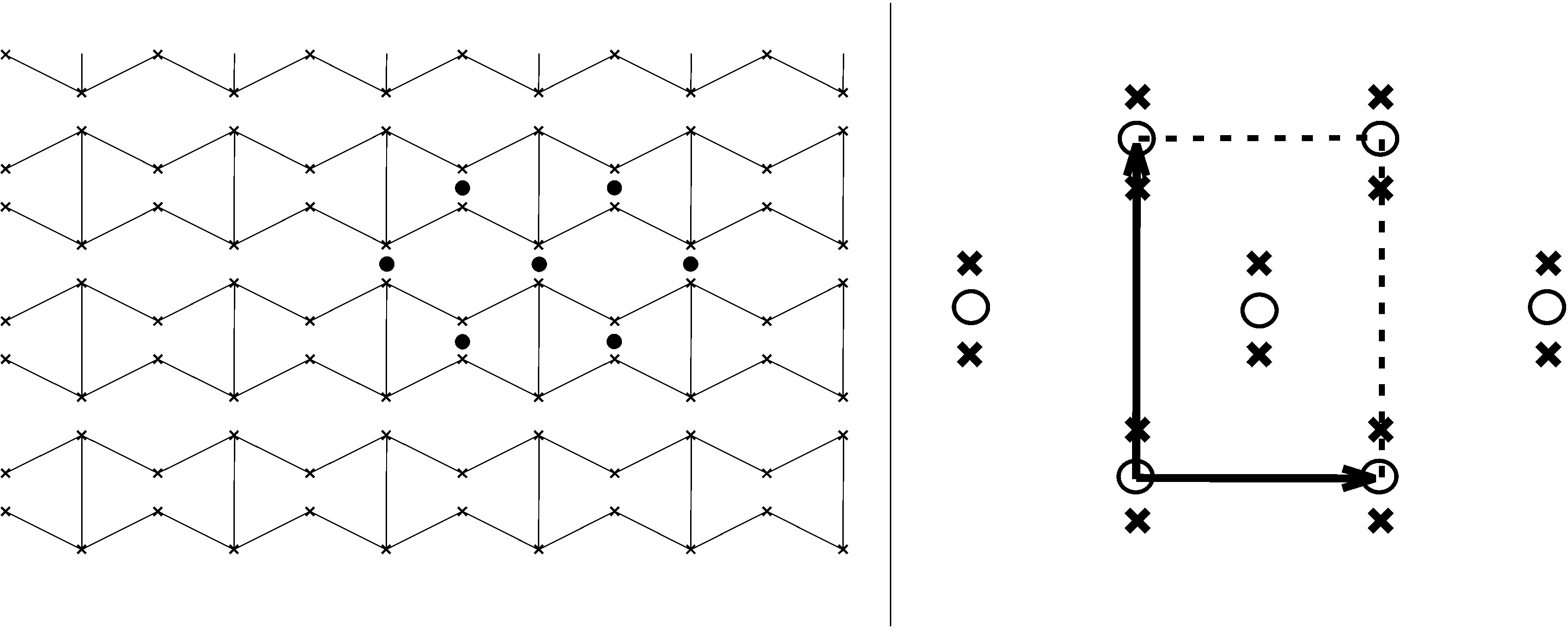}
	\caption{Left panel: Re-entrant honeycomb structure, right panel: the corresponding 2D lattice is rectangular centered.}
	\label{fig:honeycomb}
\end{figure}

To explain microscopic mechanisms underlying the property of auxeticity, various models are proposed. Two-dimensional microscopic crystalline models make an interesting class of auxetics (cf. \cite{KWW}-\cite{tret-woj}). Each model of such kind belongs to one of 2D crystal systems, therefore its auxetic properties are characterized by Poisson's ratio suitable for this system. The classic example of such structure is re-entrant honeycomb structure shown in Fig. \ref{fig:honeycomb} (cf. ref. \cite{alderson}). In Sect. \ref{sc:anisotr-oblique-rectang} we show examples of crystals of oblique and rectangular symmetry which are completely auxetic, auxetic and non-auxetic.  

\section{Anisotropy of material constants of 2D media}
\label{sc:anisotropy}
The treatment of anisotropy of Young's and shear modules and of Poisson's ratio is based on expressions derived in the companion paper \cite{japawol}. These expressions do not depend on the choice of the reference axes. However, here we prefer to use their form valid in crystalline reference axes (CRA). We denote the unit vectors parallel to the crystal axes by $\textbf{u}$, and $\textbf{v}$. We shall outline the derivation of explicit angular dependence of the elastic material constants for oblique crystal system. Suitable expressions for rectangular, quadratic and isotropic media can be obtained by reduction the number of independent and non-vanishing compliances. 

In the case of oblique 2D materials we obtained \cite{japawol}
\begin{eqnarray}
E^{-1}(\textbf{n})=s_{11}(\textbf{n}\textbf{u})^{4}+s_{22}(\textbf{n}\textbf{v})^{4}+2\left(s_{33}+s_{12}\right)(\textbf{n}\textbf{u})^{2}(\textbf{n}\textbf{v})^{2}\nonumber \\
	+2\sqrt{2}s_{13}(\textbf{n}\textbf{u})^{3}(\textbf{n}\textbf{v})+2\sqrt{2}s_{23}(\textbf{n}\textbf{u})(\textbf{n}\textbf{v})^{3},\label{eq:E-oblique} \\
-\nu(\textbf{m},\textbf{n})/E(\textbf{n})=s_{11}(\textbf{m}\textbf{u})^{2}(\textbf{n}\textbf{u})^{2}+s_{22}(\textbf{m}\textbf{v})^{2}(\textbf{n}\textbf{v})^{2}\nonumber \\
+2s_{33}(\textbf{m}\textbf{u})(\textbf{m}\textbf{v})(\textbf{n}\textbf{u})(\textbf{n}\textbf{v})
+s_{12}\left[(\textbf{m}\textbf{u})^{2}(\textbf{n}\textbf{v})^{2}+(\textbf{m}\textbf{v})^{2}(\textbf{n}\textbf{u})^{2}\right]\nonumber \\
+\sqrt{2}s_{13}\left[(\textbf{m}\textbf{u})^{2}(\textbf{n}\textbf{u})(\textbf{n}\textbf{v})+(\textbf{n}\textbf{u})^{2}(\textbf{m}\textbf{u})(\textbf{m}\textbf{v})\right]\nonumber \\
+\sqrt{2}s_{23}\left[(\textbf{m}\textbf{v})^{2}(\textbf{n}\textbf{u})(\textbf{n}\textbf{v})+(\textbf{n}\textbf{v})^{2}(\textbf{m}\textbf{u})(\textbf{m}\textbf{v})\right],
	\label{nu-oblique}
\end{eqnarray} 
\begin{eqnarray}
\left[4G(\textbf{m},\textbf{n})\right]^{-1}=s_{11}(\textbf{m}\textbf{u})^{2}(\textbf{n}\textbf{u})^{2}+s_{22}(\textbf{m}\textbf{v})^{2}(\textbf{n}\textbf{v})^{2}\nonumber \\
+s_{33}\left[(\textbf{m}\textbf{u})(\textbf{n}\textbf{v})+ (\textbf{m}\textbf{v})(\textbf{n}\textbf{u})\right]^{2}/2
+2s_{12}(\textbf{m}\textbf{u})(\textbf{n}\textbf{u})(\textbf{m}\textbf{v})(\textbf{n}\textbf{v})\nonumber \\
+\sqrt{2}\left[s_{13}(\textbf{m}\textbf{u})(\textbf{n}\textbf{u})+s_{23}(\textbf{m}\textbf{v})(\textbf{n}\textbf{v})\right]\left[(\textbf{m}\textbf{u})(\textbf{n}\textbf{v})+(\textbf{m}\textbf{v})(\textbf{n}\textbf{u})\right],
	\label{eq:shear-oblique}
\end{eqnarray}
with $s_{ij}\; (i,j=1,2,3)$ is given in the companion paper \cite{japawol}.

The vectors of stretch $\textbf{n}$ and strain $\textbf{m}$ are two mutually perpendicular vectors of the unit length. Their components fulfill three conditions
\begin{equation}
	(\textbf{n}\textbf{u})^{2}+(\textbf{n}\textbf{v})^{2}=1,\, (\textbf{m}\textbf{u})^{2}+(\textbf{m}\textbf{v})^{2}=1,\, (\textbf{n}\textbf{u})(\textbf{m}\textbf{u})+(\textbf{n}\textbf{v})(\textbf{m}\textbf{v})=0.
	\label{eq:nm-cond}
\end{equation}
This means that only one component of $\textbf{n}$ and $\textbf{m}$ is independent. In CRA the angle $\varphi$ between the vector $\textbf{u}$ and vector $\textbf{n}$ is defined by the scalar product $(\textbf{n}\textbf{u})=\cos\varphi$.  The components of $\textbf{m}$ obey the relations $(\textbf{m}\textbf{u})=m_{1}=-n_{2}=-\sin\varphi$, $(\textbf{m}\textbf{v})=m_{2}=n_{1}=\cos\varphi$. 
\subsection{Anisotropy of Poisson's ratio}
\label{nu-anisotropy} 
In CRA the function $[-\nu_{\rm{o}}(\textbf{m},\textbf{n})/E_{\rm{o}}(\textbf{n})]$ can be written in the following form
\begin{equation}
	-\nu_{\rm{o}}(\varphi)/E_{\rm{o}}(\varphi)=\left[-\nu_{\rm{o}}/E_{\rm{o}}\right]_{0}+r_{\rm{\nu}}\cos\left(4\varphi+ \psi_{\rm{\nu}}\right),
	\label{eq:nu-oblique}
\end{equation}
where 
\begin{eqnarray}
	\left[-\nu_{\rm{o}}/E_{\rm{o}}\right]_{0}=\left[\left(S_{11}+S_{22}-S_{66}\right)/2+3S_{12}\right]/4,\nonumber \\
	r_{\rm{\rm{\nu}}}=\sqrt{\left(S_{26}-S_{16}\right)^{2}+\left[S_{12}-\left(S_{11}+S_{22}-S_{66}\right)/2\right]^{2}}/4,\nonumber \\
	\tan\psi_{\rm{\rm{\nu}}}=\frac{S_{26}-S_{16}}{S_{12}-\left(S_{11}+S_{22}-S_{66}\right)/2}.
\end{eqnarray}

For rectangular, quadratic and isotropic media $S_{16}=S_{26}=0$. This means that for them $\psi_{\rm{\nu}}=0$. Adapting Eq. (\ref{eq:nu-oblique}) for oblique materials to rectangular, quadratic and isotropic materials we obtain 
\begin{equation}
-\frac{\nu_{\rm{r}}(\varphi)}{E_{\rm{r}}(\varphi)}=\left(-\frac{\nu_{\rm{r}}}{E_{\rm{r}}}\right)_{0}
 +\left[S_{12}-\left(S_{11}+S_{22}-S_{66}\right)/2\right]\cos(4\varphi)/4,
 \label{eq:crystal-nu-rect}
\end{equation}
where 
\begin{equation}
	\left(-\nu_{\rm{r}}/E_{\rm{r}}\right)_{0}=\left[\left(S_{11}+S_{22}-S_{66}\right)/2+3S_{12}\right]/4.\nonumber
\end{equation}
For quadratic media $S_{11}=S_{22}$ and 
\begin{equation}
-\nu_{\rm{q}}(\varphi)/E_{\rm{q}}(\varphi)=\left(-\nu_{\rm{q}}/E_{\rm{q}}\right)_{0}
+\left[S_{12}-\left(2S_{11}-S_{66}\right)/2\right]\cos(4\varphi)/4,
\label{eq:crystal-qu-nu}
\end{equation}
where
\begin{equation}
	\left(-\nu_{\rm{q}}/E_{\rm{q}}\right)_{0}=\left[\left(2S_{11}-S_{66}\right)/2+3S_{12}\right]/4. 
\label{-nu/E_0}
\end{equation}
For isotropic media $S_{66}=2\left(S_{11}-S_{12}\right)$ (cf. \cite{japawol}), hence form Eq. (\ref{eq:crystal-qu-nu}) we obtain 
\begin{equation}
\nu_{is}/E_{is}=-S_{12}=C_{12}/(C_{11}^{2}-C_{12}^{2}).	
\label{eq:nu-isotr}
\end{equation}
From last of Eqs. (\ref{eq:E-on-phi}) and Eq. (\ref{eq:nu-isotr}) it follows that $\nu_{is}=-S_{12}/S_{11}$. An isotropic material is mechanically stable if $S_{11}>|S_{12}|$ (cf. \cite{japawol}). Thus, in agreement with ref. \cite{wojciechowski}, two-dimensional isotropic systems can exhibit Poisson's ratio within the range $-1\leq\nu_{is} \leq +1$.
\subsection{Anisotropy of the shear modulus}
\label{g-anisotropy}
For oblique materials in the CRA Eq. (\ref{eq:shear-oblique}) yields the following expression for $\left(4G\right)^{-1}$,
\begin{equation}
	\left[4G_{\rm{o}}(\varphi)\right]^{-1}=\left(4G_{\rm{o}}\right)^{-1}_{0}+r_{\rm{G}}\cos(4\varphi+\psi_{G}),
\label{eq:G-crystal-obl}
\end{equation}
where
\begin{eqnarray}
	\left(4G_{\rm{o}}\right)^{-1}_{0}=(S_{11}+S_{22}-2S_{12}+S_{66})/8,\nonumber \\
	r_{\rm{G}}=\left[\left(S_{66}+2S_{12}-S_{11}-S_{22}\right)^{2}/4+\left(S_{26}-S_{16}\right)^{2}\right]^{1/2}/4, \nonumber\\
	\tan\psi_{G}=\frac{2\left(S_{16}-S_{26}\right)}{\left(S_{66}+2S_{12}-S_{11}-S_{22}\right)}. \nonumber
\end{eqnarray}
For rectangular, quadratic and isotropic media, $\psi_{\rm{G}}=0$ and 
\begin{equation}
\left[4G_{\rm{r}}(\varphi)\right]^{-1}=\left(4G_{\rm{o}}\right)^{-1}_{0}+\left(S_{66}+2S_{12}-S_{11}-S_{22}\right)\cos(4\varphi)/8,\nonumber
\end{equation}
and for quadratic media
\begin{equation}
\left[4G_{\rm{q}}(\varphi)\right]^{-1}=\left(4G_{\rm{q}}\right)^{-1}_{0}+\left(S_{66}+2S_{12}-2S_{11}\right)\cos(4\varphi)/8,
\nonumber
\end{equation}
where
\begin{equation}
	\left(4G_{\rm{q}}\right)^{-1}_{0}=\left(2S_{11}-2S_{12}+S_{66}\right)/8.\nonumber
\end{equation}
For isotropic media, the last formula gives the familiar result \cite{japawol}
\begin{equation}
	G^{-1}_{is}=2/\left(C_{11}-C_{12}\right)>0.
	\label{G-is}
\end{equation}
  
\subsection{Anisotropy of Young's modulus}
\label{anisotropy-Young}
Young's modulus has more complicated structure than Poisson's ratio and the shear modulus. The expressions given in Sect. \ref{sc:anisotropy} yield 
\begin{eqnarray}
E^{-1}_{\rm{o}}=S_{11}\cos^{4}\varphi+S_{11}\sin^{4}\varphi+2S_{16}\cos^{3}\varphi\sin\varphi+2S_{16}\cos\varphi\sin^{3}\varphi\nonumber \\+\left(2S_{12}+S_{66}\right)\cos^{2}\varphi \sin^{2}\varphi,\nonumber\\
E^{-1}_{\rm{r}}=S_{11}\cos^{4}\varphi+S_{22}\sin^{4}\varphi+\left(2S_{12}+S_{66}\right)\cos^{2}\varphi \sin^{2}\varphi,\nonumber\\
E^{-1}_{\rm{q}}=S_{11}\left(\cos^{4}\varphi + \sin^{4}\varphi\right)+\left(2S_{12}+S_{66}\right)\sin^{2}\varphi\cos^{2}\varphi,\nonumber\\
E^{-1}_{\rm{is}}=S_{11}.
\label{eq:E-on-phi}
\end{eqnarray}

Formulas (\ref{eq:crystal-nu-rect})-(\ref{eq:E-on-phi}) can be obtained with the help of general formulas (\ref{eq:E-oblique})-(\ref{eq:shear-oblique}) for material constants, or directly in CRA from their definitions (cf. Eqs. (33)-(35) of ref. \cite{japawol}).  
\section{Auxetic properties of 2D media}
\label{sc:auxeticity} 
\subsection{Oblique and rectangular crystals}
\label{aux-obliq}
Since $r_{\nu}^{(\rm{o})}$ is positive, Eq. (\ref{eq:nu-oblique}) implies that an oblique crystal is completely auxetic, i.e. $\nu$ is negative for all direction pairs (\textbf{m},\textbf{n}), or for all values of the angle $\varphi$, if, and only, if 
\begin{equation}
	\left(-\nu_{\rm{o}}/E_{\rm{o}}\right)_{0}>r_{\nu}^{(\rm{o})}>0. 
	\label{ineq:obl-complete-aux}
\end{equation}
If 
\begin{equation}
	\left(-\nu_{\rm{o}}/E_{\rm{o}}\right)_{0}<0,\; |\left(-\nu_{\rm{o}}/E_{\rm{o}}\right)_{0}|>r_{\nu}^{(\rm{o})}, 
	\label{ineq:obl-complete-nonaux}
\end{equation}
a crystal is non-auxtetic, i.e. $\nu(\textbf{n},\textbf{m})$ is positive for all direction pairs $(\textbf{m},\textbf{n})$ or for all angles $\varphi$. 

If 
\begin{equation}
	|\left(-\nu_{\rm{o}}/E_{\rm{o}}\right)_{0}|<r_{\nu}, 
	\label{ineq:obl-auxetic}
\end{equation}
a crystal is auxetic, i.e. there can be found auxetic direction pairs (\textbf{m},\textbf{n}) (angles $\varphi$) for which the right hand side of Eq. (\ref{eq:nu-oblique}) is positive, and non-auxetic direction pairs for which it is negative. 

Unlike in the case of 2D crystals of oblique symmetry, for crystals of rectangular symmetry, the parameter $r_{\nu}^{(q)}$ can be negative. 
Anisotropy of elastic material constants and auxetic properties of quadratic crystals will be studied in the next section. 
\subsection{Auxetic properties of quadratic crystals}
\label{sc:quadratic-anisotropy-auxeticity}
In the case of quadratic crystals, one may use eigenvalues $c_{J}$, $c_{L}$, and $c_{M}$ in place of three elastic constants $C_{11}$, $C_{12}$, and $C_{66}$, but the more reasonable choice is the use of three parameters, which are the analogs of the parameters introduced by Every \cite{every}, namely, $s_{1}=\left(C_{11}+C_{66}\right)$ and two dimensionless parameters
\begin{equation}
	s_{2}=\left(C_{11}-C_{66}\right)/s_{1}, \; s_{3}=K/s_{1},
\label{pasz-variables}
\end{equation}
where $K=\left(C_{11}-C_{12}-C_{66}\right)/2$ is the elastic anisotropy parameter. 
One may express the elastic constants $C_{IJ}\, (I,J=1,2,6)$ by $s_{1},\,s_{2}$ and $s_{3}$ parameters, namely
\begin{equation}
	C_{11}=s_{1}\left(1+s_{2}\right)/2,\; C_{12}=s_{1}\left(3s_{2}/2-s_{3}-1/2\right), \; C_{66}=s_{1}\left(1-s_{2}\right).  
\label{eq:Cbys}
\end{equation}
These parameters parametrize phase velocities of quadratic crystals in a natural way (cf. Appendix \ref{sc:phase-vel}).

\begin{figure}[htpb]
	\centering
		\includegraphics[width=11.5cm, bb=125 296 498 600,clip]{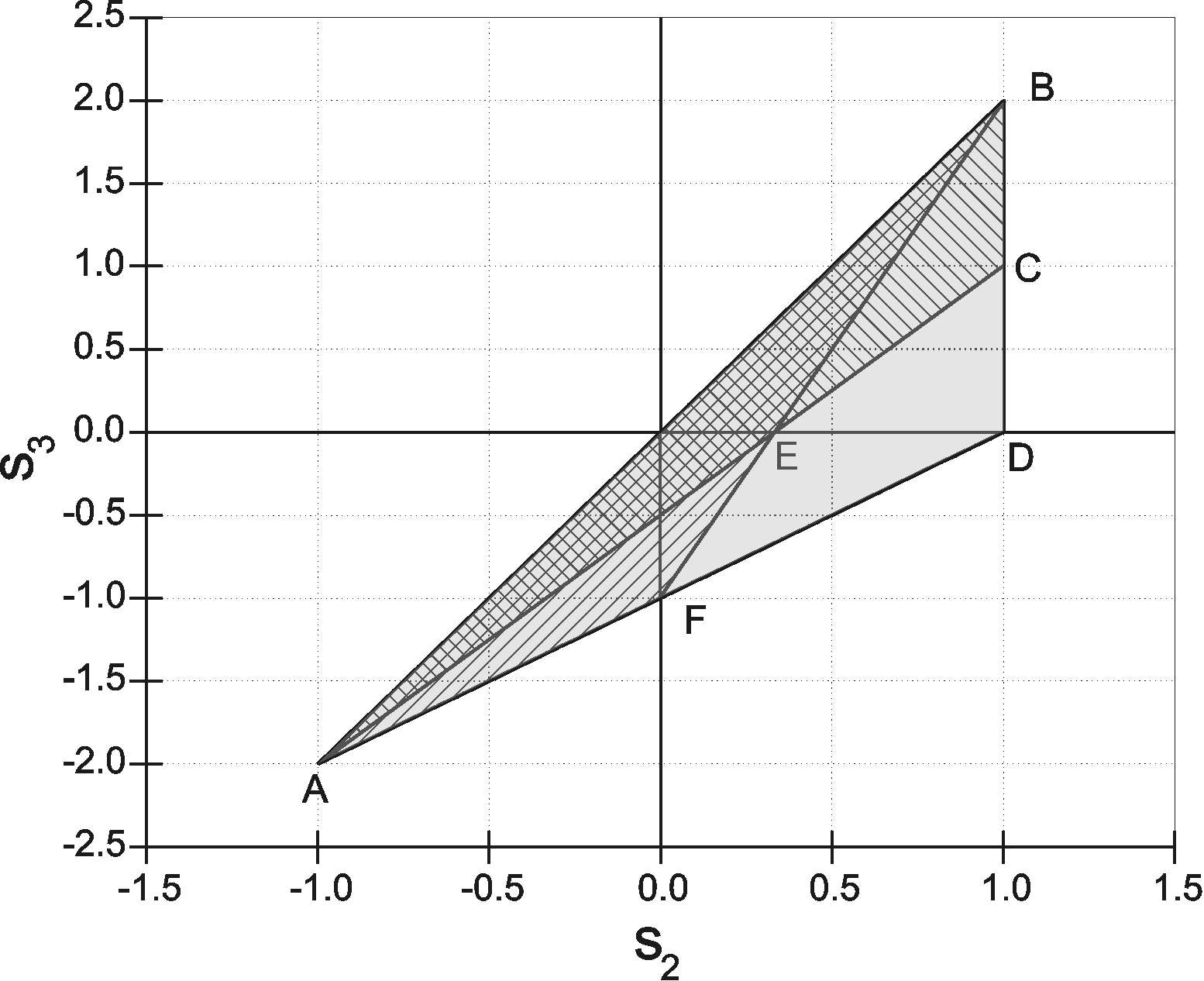}
	\caption{Triangle ABD of stability (ST) of 2D crystals of quadratic symmetry. Points $\left(s_{1}, s_{2}, s_{3}\right)$ characterizing mechanically stable quadratic materials belong to the prism ($s_{1}>0$), the base of which is the stability triangle. Points of the interval $0<s_{2}<1$ represent isotropic media.}
	\label{fig:Trojstab}
\end{figure}

In terms of parameters $s_{1}\;s_{2}$, and $s_{3}$ a quadratic material is stable, if $s_{1}>0$ and $s_{2}$, $s_{3}$ belong to the \emph{stability triangle} (ST for short) in the plane ($s_{2}$, $s_{3}$) (cf. Fig. \ref{fig:Trojstab}) defined by three inequalities 
\begin{equation}
	1>s_{2},\;  \left(1-s_{2}+s_{3}\right)>0,\; \left(s_{2}-s_{3}/2\right)>0. 
\label{stability-cond}
\end{equation}
Sides of ST are defined by lines $AB$ ($s_{3}=2s_{2}$), $BD$ ($s_{2}=1$) and $AD$ ($s_{3}=s_{2}-1$). 

The set of inequalities 
\begin{equation}
	\left(-\nu_{q}/E_{q}\right)_{0}>0,\; \left(-\nu_{q}/E_{q}\right)_{0}>|r_{\nu}^{(q)}|, 
\label{ineq:compl-aux}
\end{equation}
is equivalent to inequalities shown in the I and II row of the Table \ref{table:compliances-vs-every}, and defines the region of ST in form of the triangle $AEB$ representing crystals of quadratic symmetry which are completely auxetic. 
   
A crystal of quadratic symmetry is non-auxetic when 
\begin{equation}
	\left(-\nu_{q}/E_{q}\right)_{0}<0,\; |-\left(\nu_{q}/E_{q}\right)_{0}|>|r_{\nu}|.  
\label{ineq:non-aux}
\end{equation}
The inequalities (\ref{ineq:non-aux}) are equivalent to inequalities IV and V of the Table \ref{table:compliances-vs-every}, and circumscribe the quadrilateral region $FECD$ of ST. 

A crystal of quadratic symmetry is auxetic, provided that 
\begin{equation}
	\left|\left(-4\nu_{q}/E_{q}\right)_{0}\right|<\left|r_{\nu}^{(q)}\right|.
\label{ineq:quadr-aux}
\end{equation}
If $r_{\nu}^{(q)}>0$, then the inequality (\ref{ineq:quadr-aux}) is equivalent to two pairs of conditions 
\begin{eqnarray}
	\left(-4\nu_{q}/E_{q}\right)_{0}>0, r_{\nu}^{(q)}>\left(-4\nu_{q}/E_{q}\right)_{0}, \nonumber \\		   				      
	\left(-4\nu_{q}/E_{q}\right)_{0}<0, r_{\nu}^{(q)}>-\left(-4\nu_{q}/E_{q}\right)_{0}, 
\label{ineq:top-auxetic}
\end{eqnarray}
which are equivalent to inequalities I and V defining the triangle $EBC$. 

If $r_{\nu}^{(0)}<0$, then in agreement with inequalities I and IV of Table \ref{table:compliances-vs-every}
\begin{eqnarray}
	\left(-4\nu_{q}/E_{q}\right)_{0}>0, \left(-4\nu_{q}/E_{q}\right)_{0}<-r_{\nu}^{(q)}, \nonumber \\		   				      
	\left(-4\nu_{q}/E_{q}\right)_{0}<0, -\left(-4\nu_{q}/E_{q}\right)_{0}<-r_{\nu}^{(q)}, 
\label{ineq:bottom-auxetic}
\end{eqnarray}
bound the triangle $AFE$. 

We shall mention that points of the interval $s_{3}=0,\; 0<s_{2}<1/3$ represent isotropic completely auxetic 2D materials, whereas points of the interval $s_{3}=0,\; 1/3<s_{2}<1$ represents isotropic non-auxetic 2D elastic media. 
\begin{table}
\begin{tabular}
{ll|c|c|c}
&No. &  Compliances & Every's parameters & region of ST\\ \hline
&I & $S_{12}>0$ & $\frac{3}{2}s_{2}-s_{3}-\frac{1}{2}<0$ & above AC \\ \hline
&II & $S_{11}+S_{12}-S_{66}/2>0$ & $s_{3}-3s_{2}+1>0$ & above BF \\ \hline
&III & $r_{\nu}^{(q)}>0$ & $s_{3}>0$ & upper part of ST \\ \hline
&IV & $S_{12}<0$ & $\frac{3}{2}s_{2}-s_{3}-\frac{1}{2}>0$ & below AC \\ \hline
&V & $S_{11}+S_{12}-S_{66}/2<0$ & $s_{3}-3s_{2}+1<0$ & below BF \\ \hline
&VI & $r_{\nu}^{(q)}<0$ & $s_{3}<0$ & lower part of ST  \\
\end{tabular}
\caption{}
\label{table:compliances-vs-every}
\end{table}
\section{Anisotropy of elastic material constants of 2D crystals}
\label{sc:anisotropy_general}
\subsection{Anisotropy of elastic material constants of crystals of oblique and rectangular symmetry}
\label{sc:anisotr-oblique-rectang}
\begin{figure}[htpb]
	\centering
		\includegraphics[width=11.5cm, bb=50 50 410 302 , clip]{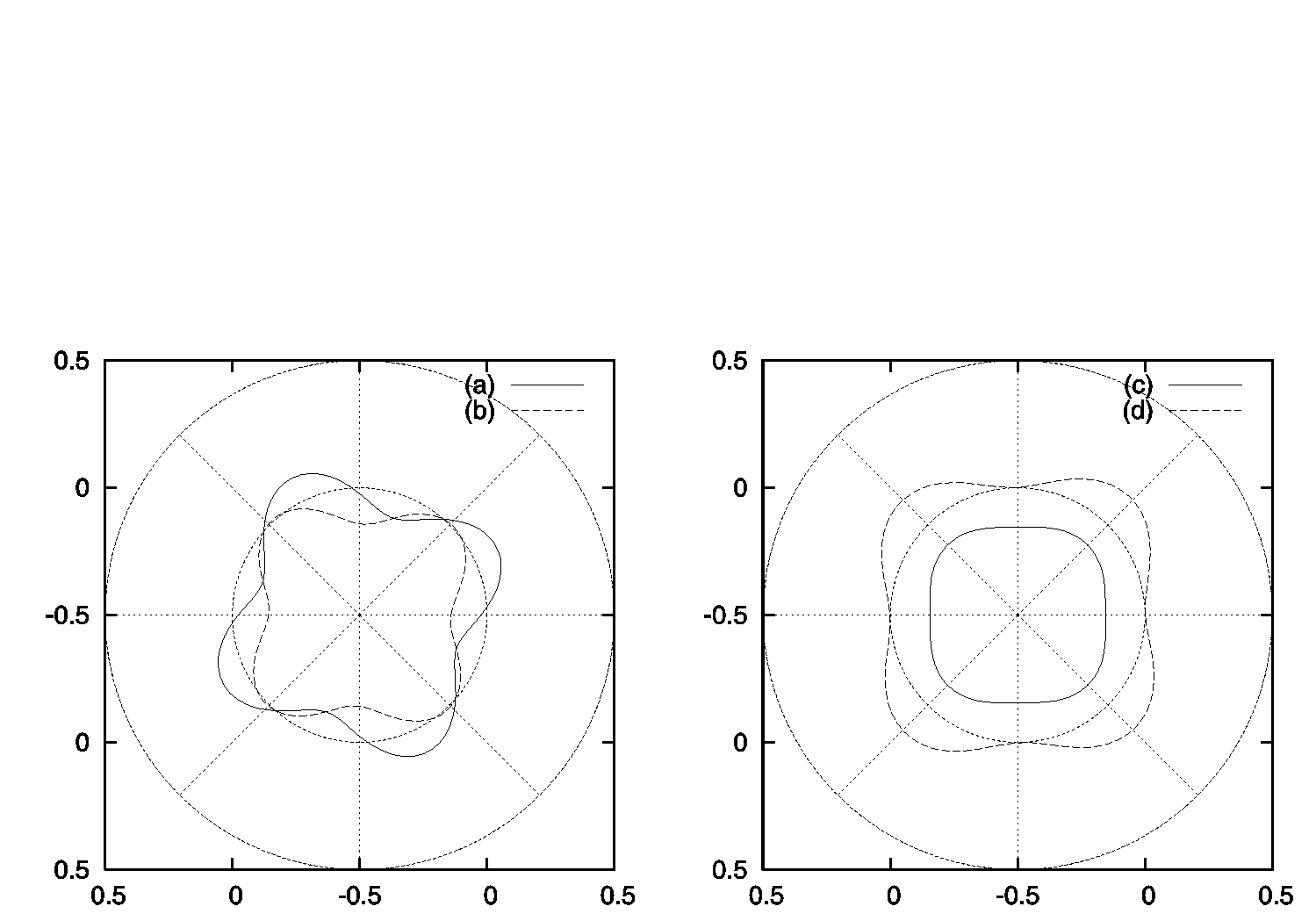}
\caption{Dependence of $\nu({\bf m},{\bf n})/E({\bf n})$ of crystals of oblique symmetry on angle $\varphi$ for materials with compliances collected in Table \ref{table:compl-obl}. (a) and (b) -- auxetic,  (c) -- completely auxetic, and (d) -- non-auxetic behavior.}
\label{fig:poisson_oblique}
\end{figure}

\begin{figure}[htpb]
	\centering
		\includegraphics[width=11.5cm, bb=50 50 410 302 , clip]{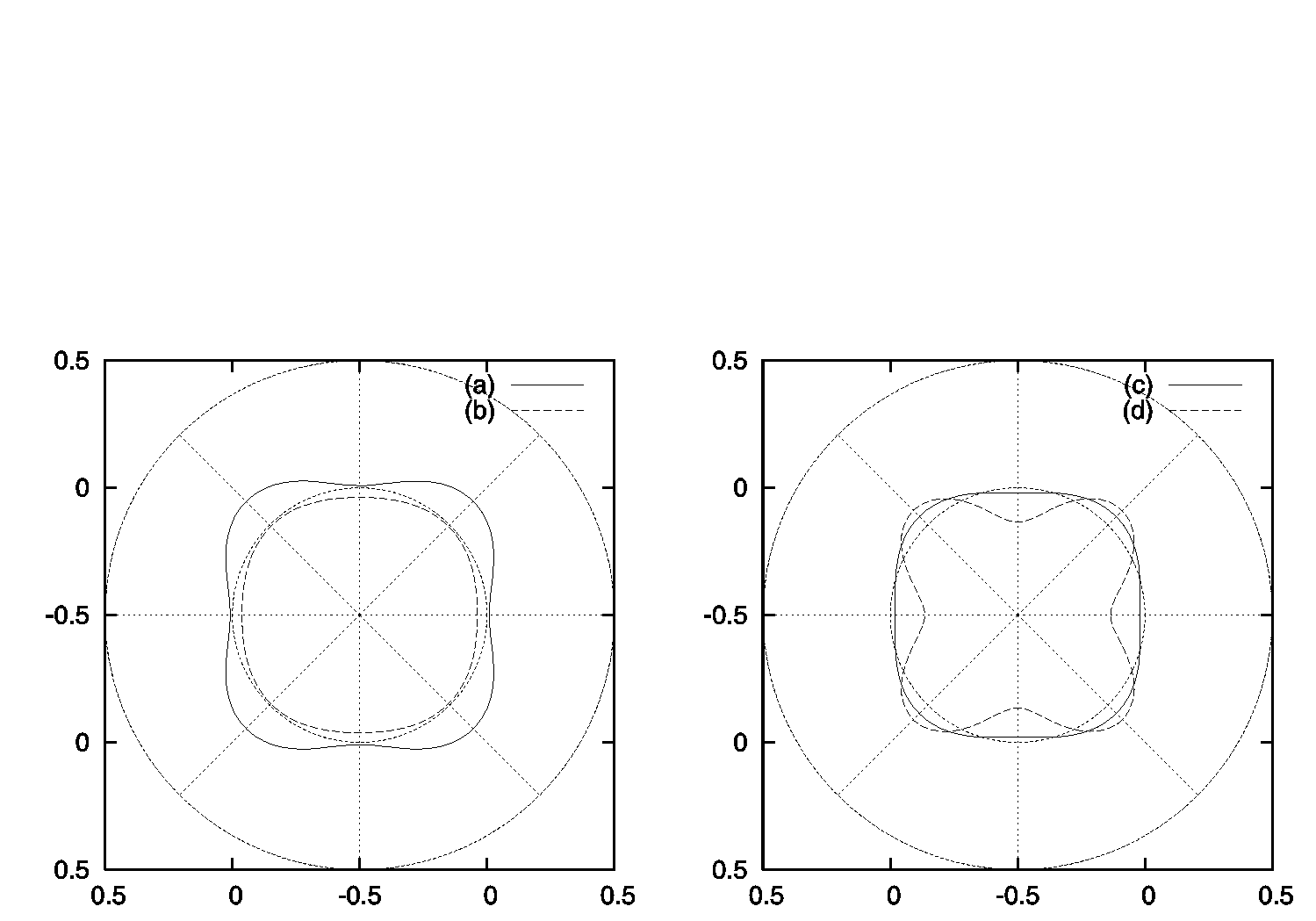}
\caption{Dependence of $\nu({\bf m},{\bf n})/E({\bf n})$  of crystals of rectangular symmetry on the angle $\varphi$ for materials listed in Table \ref{table:comp-rectangular}. (a) -- non-auxetic,  (b) -- completely auxetic, (c) and (d) -- auxetic behavior.}
\label{fig:poisson_rect}
\end{figure}

In the case of crystals of oblique and rectangular symmetry we restrict the discussion of the anisotropy to Poisson's coefficient. Using compliances collected in the Appendix \ref{sc:appendix2} we draw polar plots for Poisson's coefficient of oblique and rectangular media. They are shown in Figs. \ref{fig:poisson_oblique} and \ref{fig:poisson_rect}. 

Inspecting these figures we conclude that there exits completely auxetic, auxetic and non-auxetic crystals of oblique and rectangular symmetry. 
\subsection{Anisotropy of elastic material constants of crystals of quadratic symmetry}
\label{anisotr-quadratic}
\begin{figure}[htpb]
	\centering
		\includegraphics[width=11.5cm, bb=50 50 410 302, clip]{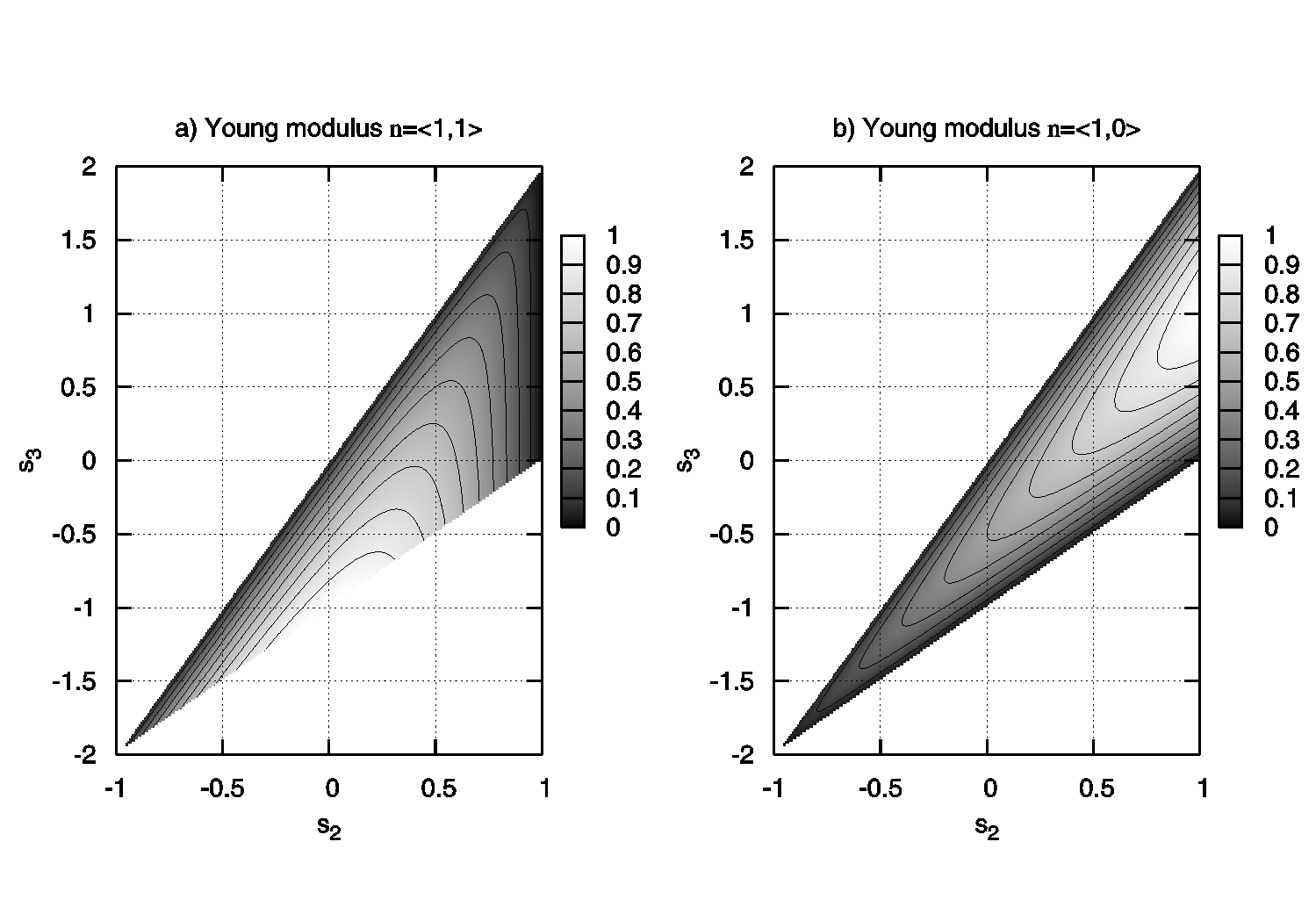}
	\caption{Quadratic crystals. Contour map of dimensionless Young modulus $e_{q}\left(s_{2},s_{3};\mathbf{n}\right)$ for \newline $\textbf{n}=<1,1>$ and $\textbf{n}=<1,0>$}
	\label{fig:young-modulus}
\end{figure}

We shall characterize the response of the elastic quadratic material to uniaxial tension in a direction $\textbf{n}\; (\textbf{n}\textbf{n}=1)$ by the dimensionless Young's modulus $e(\textbf{n})=E(\textbf{n})/s_{1}$ (cf. ref. \cite{pasz-wol} and \cite{japawol})
\begin{equation}
	e^{-1}(\textbf{n})=(\textbf{n}\otimes\textbf{n})\cdot\textbf{S}'\cdot(\textbf{n}\otimes\textbf{n})=n_{i}n_{j}S_{ijkl}'n_{k}n_{l},
\end{equation}
where $S_{ijkl}'=s_{1}S_{ijkl}$. Repetition of a suffix in a product of tensors implies the usual summation with the respect to that suffix over the values 1,2.
In Fig. \ref{fig:young-modulus} we present the dependence of dimensionless Young modulus $e=E/s_{1}$ on the parameters $s_{2}$, $s_{3}$ for the same  characteristic directions as above. 

Similarly, the response to the shear stress will be characterized by the dimensionless shear modulus $g(\textbf{n},\textbf{m})=G(\textbf{n},\textbf{m})/s_{1}$, ($\textbf{m}\textbf{m}=1,\;\textbf{n}\textbf{m}=0$)
\begin{equation}
\frac{1}{4g(\textbf{m},\textbf{n})}=(\textbf{m}\otimes\textbf{n})\cdot S'\cdot(\textbf{m}\otimes\textbf{n})=m_{i}n_{j}S_{ijkl}'m_{k}n_{l}.  
\label{eq:shear}
\end{equation}

\begin{figure}[htpb]
	\centering
		\includegraphics[width=11.5cm, bb=50 50 410 302, clip]{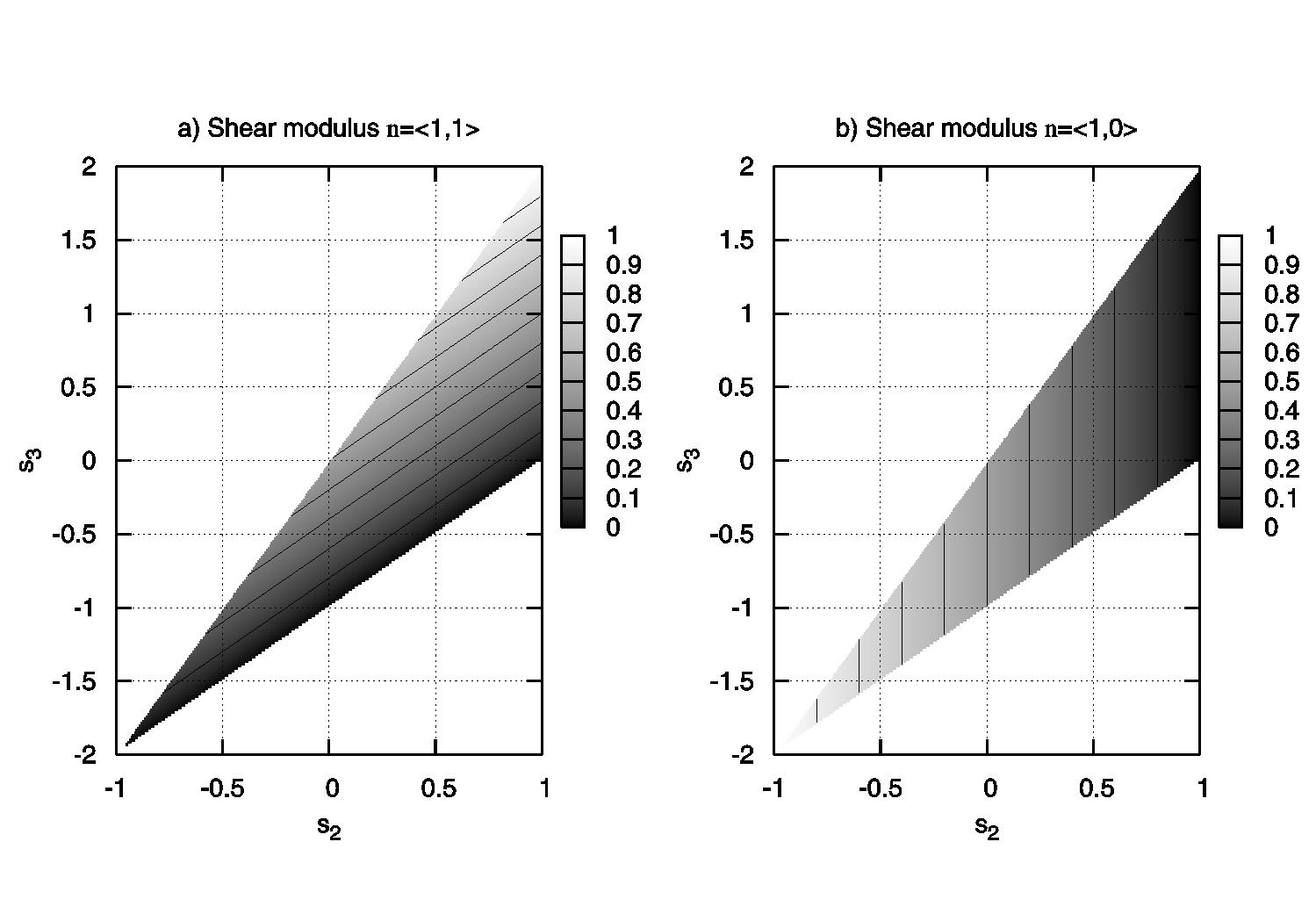}
	\caption{Quadratic crystals. Contour map of dimensionless shear modulus $g_{q}\left(s_{2},\,s_{3};\,\mathbf{m},\,\mathbf{n}\right)$ for $\textbf{n}=<1,1>$, $\textbf{m}=<-1,1>$, and $\textbf{n}=<1,0>,\; \textbf{m}=<0,1>$}
	\label{fig:shear-modulus}
\end{figure}

The dependence of $g_{q}$ on $s_{2}$ and $s_{3}$ for the same characteristic directions as in the case of Young's modulus is shown in Fig. \ref{fig:shear-modulus}.

The dimensionless Poisson's ratio $\nu(\textbf{n},\textbf{m})$ is the ratio of the lateral contraction along direction $\textbf{m}$ to the longitudinal extension along  direction $\textbf{n}$ perpendicular to $\textbf{m}$
\begin{equation}
\nu(\textbf{m},\textbf{n})=-e(\textbf{n})(\textbf{m}\otimes\textbf{m})\cdot S'\cdot(\textbf{n}\otimes\textbf{n}) = -\frac{m_{i}m_{j}S_{ijkl}'n_{k}n_{l}}{n_{t}n_{u}S_{tuvw}'n_{v}n_{w}}. 
\label{eq:poisson}
\end{equation}

\begin{figure}[htpb]
	\centering
		\includegraphics[width=11.5cm, bb=50 50 410 302, clip]{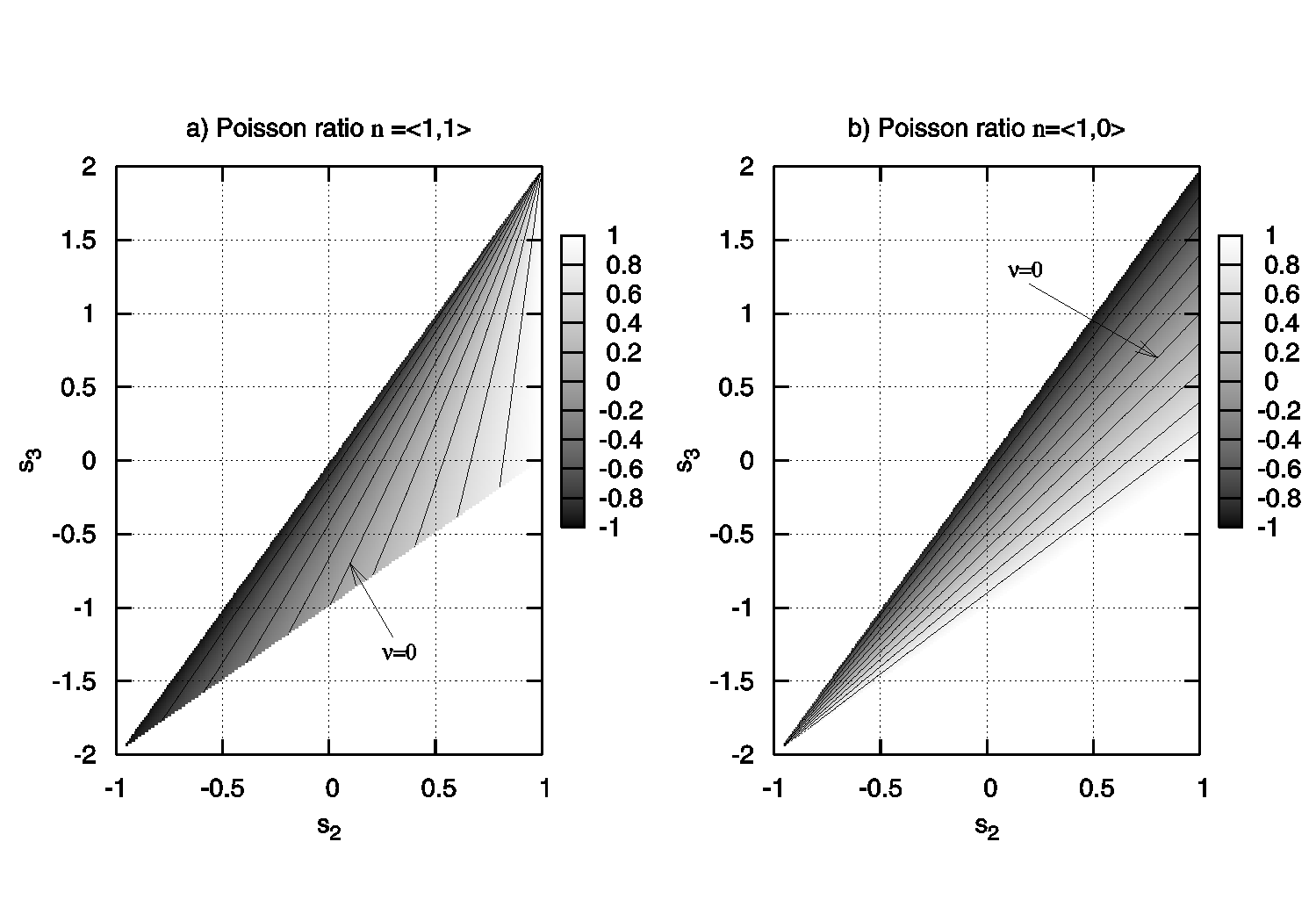}
	\caption{Quadratic crystals. Contour map of dimensionless Poisson's ratio $\nu_{q}\left(s_{2},s_{3};\mathbf{m},\mathbf{n}\right)$ for $\textbf{n}=<1,1>$, $\textbf{m}=<-1,1>$, and $\textbf{n}=<1,0>,\; \textbf{n}=<0,1>$}
	\label{fig:poisson-ratio}
\end{figure}

For quadratic crystals the dependence of dimensionless Poisson's ratio $\nu/s_{1}$ on parameters $s_{2}$, $s_{3}$ for two characteristic directions $\textbf{n}=<1,1>$ and $\textbf{n}=<1,0>$ is shown in Fig. \ref{fig:poisson-ratio}. Inspecting these figures we note that in the case of quadratic materials ${\nu}$ obeys the inequality established for isotropic media in Sect. \ref{nu-anisotropy}.

\begin{figure}[htpb]
	\centering
		\includegraphics[width=11.5cm, bb=50 50 410 302, clip]{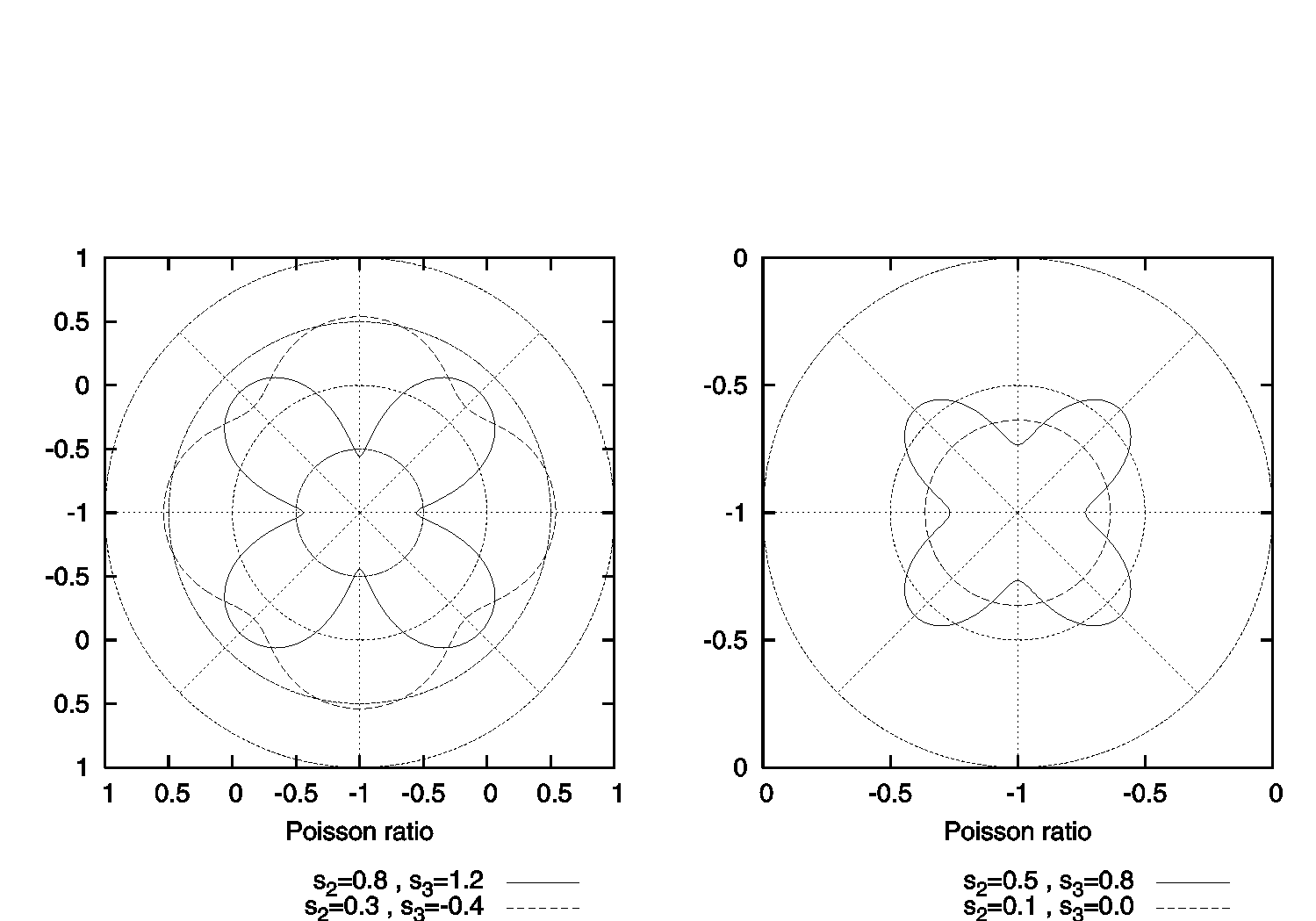}
	\caption{Dependence of $\nu_q({\bf m},{\bf n})/E_q({\bf n})$ on the angle $\varphi$ for several points belonging to characteristic regions of ST for quadratic crystals. 
	Left diagram -- full line represents auxetic behavior, broken line -- non-auxetic. Right diagram illustrates completely auxetic behavior of anisotropic (full line) and isotropic material (broken line).}
	\label{fig:poisson_polar}
\end{figure}

Finally, the results of our calculation of angular dependence of Poisson's ratio for points belonging to the characteristic regions of ST are  presented in Fig. \ref{fig:poisson_polar}. The solid line represents complete auxetics (the region \emph{AEBG} in Fig. \ref{fig:Trojstab}). The dotted line represents complete non-auxetics (the region \emph{FECD}) and the broken line -- non-complete auxetics (the region \emph{AFE} and \emph{ECB}). For them Poisson's ratio has to be anisotropic. The point $s_{2}=0.1,\; s_{3}=0$ represents complete isotropic auxetics.  
All these contours have fourfold symmetry in agreement with formula (\ref{eq:crystal-qu-nu}), but their orientations depend on coordinates $s_{2},s_{3}$.
\appendix
\section{Phase velocities for quadratic materials}
\label{sc:phase-vel}
Suppose that an acoustic wave propagates in a quadratic medium with the density $\rho$ in the direction $\textbf{n}$. In the reference frame related to the symmetry axes of a quadratic lattice, elements $\Gamma_{ij}=n_{p}C_{ipjq}n_{q}$ of Christoffel's matrix $\Gamma$ are given by \cite{japawol}
\begin{eqnarray}
\Gamma_{11}=\left(C_{11}n_{1}^{2}+C_{66}n_{2}^{2}\right)=C_{2}n_{1}^{2}+C_{66},\nonumber \\
\Gamma_{22}=\left(C_{11}n_{2}^{2}+C_{66}n_{1}^{2}\right)=C_{2}n_{2}^{2}+C_{66} ,\nonumber \\ 
	\Gamma_{12}=\Gamma_{21}=\left(C_{12}+C_{66}\right)n_{1}n_{2}=\left(C_{2}-K\right)n_{1}n_{2}.
	\label{eq:christof-matrix} 
\end{eqnarray}
$C_{1}$, $C_{2}$, and the anisotropy parameter $K$ are the analogues of parameters introduced by Every \cite{every}
\begin{equation}
	C_{1}\equiv \left(C_{11}+C_{66}\right),\;C_{2}\equiv \left(C_{11}-C_{66}\right),\;K\equiv \left(C_{11}-C_{12}-2C_{66}\right).
\label{eq:every-parameters}
\end{equation}

The eigenvalues of $\Gamma$ are equal to $\rho v_{j}\, (j=\pm 1)$, where $j$ enumerates two phase velocities. Denote by $T$ the trace of Christoffel's matrix. For crystals of quadratic symmetry $T=Tr\Gamma=\left(C_{11}+C_{66}\right)$. Upon making the replacement $2\rho v^{2}=S+T$ \cite{every}, one arrives at the following eigenvalue problem equation for S 
\begin{equation}
	\left|\Lambda-SI_{2}\right| =\left|
\begin{array}
[c]{cc}%
2C_{2}\left(n_{1}^{2}-1/2\right)-S & \left(C_{2}-K\right)
n_{1}n_{2}\\
\left(C_{2}-K\right)n_{1}n_{2} & 2C_{2}\left(n_{2}^{2}-1/2\right)-S
\end{array}
\right|=0.
\label{eq:eigen-problem}
\end{equation}
On expanding the determinant in Eq. (\ref{eq:eigen-problem}) one obtains a quadratic equation, whose solutions reads
\begin{equation}
	v_{\pm}^{2}=\frac{1}{2\rho}\left[C_{1}\pm 2\sqrt{\left(C_{2}-K\right)^{2}n_{1}^{2}n_{2}^{2}-C_{2}^{2}\left(n_{1}^{2}-1/2\right)\left(n_{2}^{2}-1/2\right)}\right].
\label{eq:phase-vel}
\end{equation}

For isotropic media, $C_{66}=\left(C_{11}-C_{12}\right)/2$ and $S_{66}=2\left(S_{11}-S_{12}\right)$, hence for them $K=0$. Thus, the condition for elastic isotropy is that $K=0$. When this is satisfied, one obtains the phase velocities for 2D isotropic media \cite{japawol} 
\begin{equation}
	v_{+}\equiv v_{l}=\sqrt{C_{11}/\rho}, \; v_{-}\equiv v_{t}=\sqrt{\left(C_{11}-C_{12}\right)/2\rho}.\nonumber
\end{equation} 

\section{Examples of compliances for crystals of oblique and rectangular symmetry}
\label{sc:appendix2}
We have written a computer program which finds numerical values of compliances obeying the mechanical stability conditions described in ref. \cite{pasz-wol}. Several such sets for oblique and rectangular materials are collected in tables \ref{table:compl-obl} and \ref{table:comp-rectangular}

\begin{table}
\begin{tabular}
{l|l|l|l|l|l|l|c}\hline
&$S_{11}$ & $S_{12}$ & $S_{22}$ & $S_{16}$ & $S_{26}$ & $S_{66}$& aux. propert.\\ \hline
(a)&0.105    &  0.024   & 0.372    & -0.071   &  0.311   & 0.626  &  auxetic \\
(b)&0.037    & -0.006   & 0.633    &  0.038   & -0.015   & 0.090  &  auxetic \\
(c)&0.178    &  0.111   & 0.823    & -0.057   & -0.053   & 0.601  & compl. auxetic  \\
(d)&0.037    & -0.001   & 0.321    &  0.033   &  0.0004  & 0.910  &  non-auxetic
\end{tabular}
\caption{Compliances for crystals of oblique symmetry. They represent materials with all types auxetic properties.}
\label{table:compl-obl}
\end{table}

\begin{table}
\begin{tabular}
{l|l|l|l|l|c}\hline
&$S_{11}$ & $S_{12}$ & $S_{22}$ & $S_{66}$&aux. propert.\\\hline
(a)&0.077     & -0.008   & 0.215    & 0.798   &non-auxetic\\
(b)&0.100     & 0.007   & 0.315    & 0.277   &compl. auxetic\\
(c)&0.030     &  0.019   & 0.672    & 0.928   &auxetic\\
(d)&0.100     & -0.080   & 0.971    & 0.368   &auxetic
\end{tabular}
\caption{Compliances for crystals of rectangular symmetry. They represent materials with all types of auxetic properties.}
\label{table:comp-rectangular}
\end{table}

\begin{figure}[htpb]
	\centering
		\includegraphics[width=11.5cm, bb=140 50 320 302, clip]{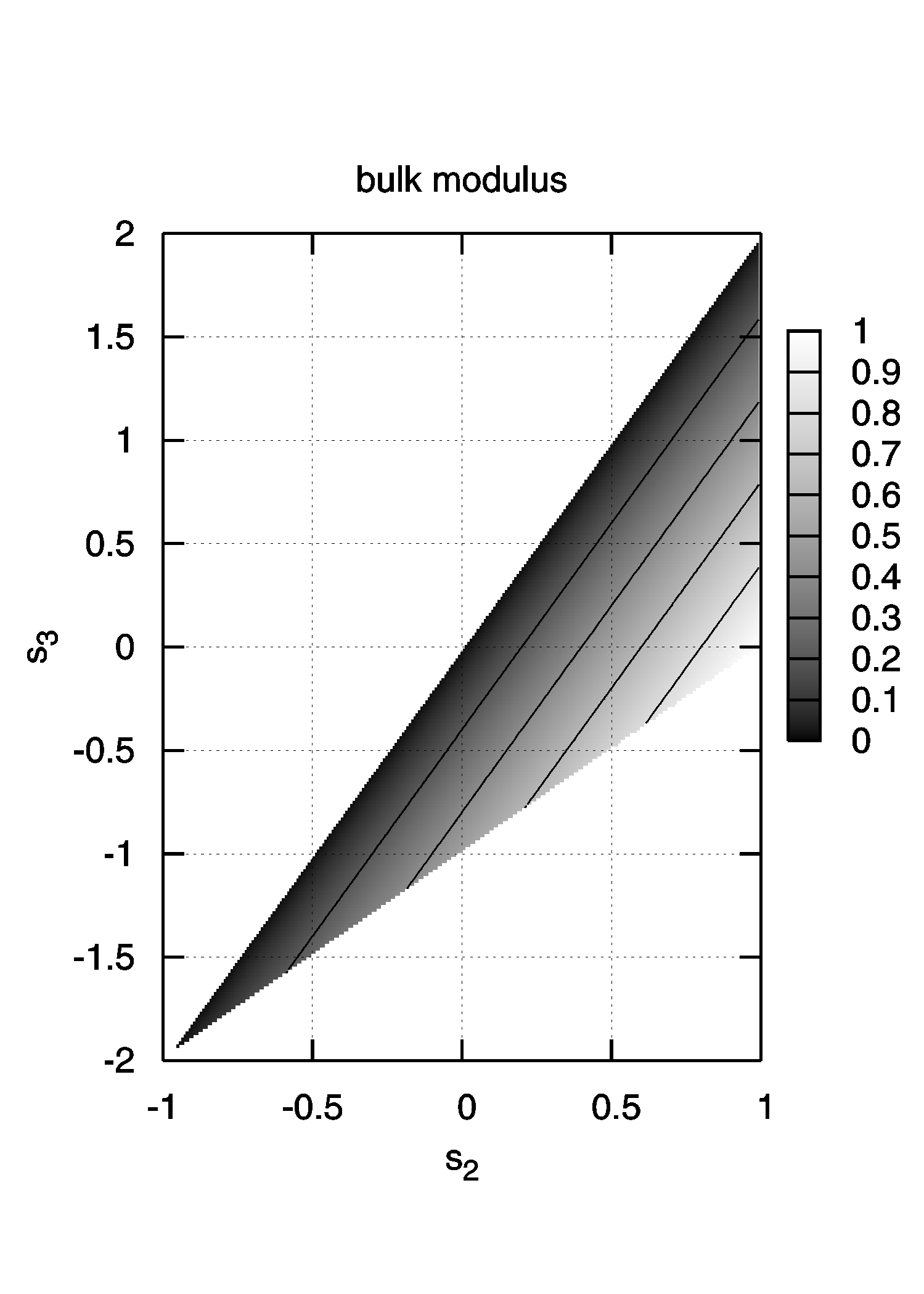}
	\caption{Dependence of dimensionless bulk modulus $\kappa'$ on $s_{2}$ and $s_{3}$ for quadratic crystals.  $\kappa'$ vanishes on the upper side of ST}
	\label{fig:kappa}
\end{figure}

These values are used for calculating angular dependence of Poisson's coefficient for oblique and rectangular materials (cf. Sect. \ref{sc:anisotr-oblique-rectang}).

\section{Instabilities on the borders of ST}
\label{sc:instabilities}
Inspecting Figs. \ref{fig:young-modulus} and \ref{fig:shear-modulus} one notice that for considered directions $\bf{n}$ and $\bf{m}$ Young's modulus and the shear modulus $G(\bf{m},\bf{n})$ vanish on various borders of the stability triangle (cf. Table \ref{table:instabilities}). The bulk modulus $\kappa$ is defined as \cite{rychlewski}
\begin{equation}
	\kappa^{-1}=I_{2}\cdot S\cdot I_{2}=S_{iijj}=S_{11}+2S_{12}+S_{22}.
\label{eq:bulk}	
\end{equation}
For quadratic crystals $\kappa=\left[2\left(S_{11}+S_{12}\right)\right]^{-1}=\left(C_{11}+C_{12}\right)/2=c_{J}/2=s_{1}\left(2s_{2}-s_{3}\right)/2=s_{1}\kappa'$ vanishes on upper side of ST (cf. also Fig. \ref{fig:kappa}). Vanishing of modules marks phase transitions. The vanishing the shear modulus is accompanied by soft modes. For 3D cubic crystals we discussed all such instabilities and possible low symmetry phases in ref. \cite{papruz}. For such crystals instability related to $\kappa=0$ leads to the discontinuous isomorphic phase transition.

\begin{table}
\begin{tabular}
{l|c|c}\hline
$E$ & $\bf{n}=\left\langle 1,1\right\rangle$ ; $BF$ & $\bf{n}=\left\langle 1,0 \right\rangle$; $AB$, $AD$ \\ \hline
$G$ & $\bf{n}=\left\langle1,1\right\rangle$, $\bf{m}=\left\langle-1,1\right\rangle$; $AD$ & $\bf{n}=\left\langle 1,0\right\rangle$, $\bf{m}=\left\langle 0,1\right\rangle$; $BD$ 
\end{tabular}
\caption{Instabilities of modules for various directions. Sides of ST on which instabilities occur are indicated.}
\label{table:instabilities}
\end{table}

\end{document}